\documentclass[10pt]{article}
\usepackage[dvips]{color}
\usepackage{epsfig}
\usepackage{epsf}
\usepackage{amsmath}
\usepackage{graphicx}
\usepackage{float}

\begin{document}
\setcounter{page}{1}

\pagestyle{plain} \vspace{1cm}
\begin{center}
{\Large \bf Constraints on a scalar-tensor model with Gauss-Bonnet coupling from SN Ia and BAO observations}\\
\small \vspace{1cm} {\bf S. Bellucci $^{a}$
\footnote{Stefano.Bellucci@lnf.infn.it}}, {\bf A. Banijamali  $^{b}$
\footnote{a.banijamali@nit.ac.ir}}, {\bf B. Fazlpour
$^{c}$ \footnote{b.fazlpour@umz.ac.ir}} and {\bf M. Solbi $^{b}$}\\
\vspace{0.5cm} $^{a}$ {\it INFN - Laboratori Nazionali di Frascati,
1-00044, Frascati (Rome), Italy\\} \vspace{0.5cm} $^{b}$ {\it
Department of Basic Sciences, Babol Noshirvani University of
Technology, Babol, Iran\\} \vspace{0.5cm} $^{c}$ {\it Department of
Physics, Babol Branch, Islamic Azad University, Babol, Iran\\}
\end{center}
\vspace{1.5cm}
\begin{abstract}
In the present work, the observational consequences of a subclass of
of the Horndeski theory have been investigated. In this theory a
scalar field (tachyon field) is non-minimally coupled to the
Gauss-Bonnet invariant through an arbitrary function of the scalar
field. By considering a spatially flat FRW universe, the free
parameters of the model have been constrained using a joint analysis
from observational data of the Type Ia supernovae and Baryon
Acoustic Oscillations measurements. The best fit values obtained
from these datasets are then used to reconstruct the equation of
state parameter of the scalar field. The results show the phantom,
quintessence and phantom divide line crossing behavior of the
equation of state and also cosmological viability of the model.
\end{abstract}

\newpage
\section{Introduction}
The current accelerated expansion of the universe is one of the
great problems of modern cosmology. This acceleration was first
suggested by Type Ia supernovae (SN Ia) surveys
\cite{Riess:1998,Perlmutter:1999} and then by measurements of the
cosmic microwave background (CMB) \cite{Komatsu:2011,Ade:2014}, the
Hubble constant \cite{Riess:2009}, Baryon Acoustic Oscillations
(BAO) \cite{Lampeitl:2009} and more measurements of Type Ia
supernovae \cite{Kowalski:2008}. Although observational cosmology
confirms the acceleration of the universe, explaining this issue from
theoretical point of view is a big challenge. The simplest way to
obtain an accelerated universe is adding a cosmological constant to
the standard cosmological model. However, a cosmological constant
suffers from the fine-tuning problem, that is due to its extremely small
observed value compared to predictions from theoretical
considerations \cite{Martin:2012}. As a result one can follow two
ways to explain the late-time behavior of the universe: modifying
general relativity at large scale \cite{Capozziello:2011} or
introducing a new content in the universe such as canonical scalar
field, phantom scalar, both scalars, vector fields etc., that is
introducing the concept of dark
energy \cite{Bassett:2006,Copeland:2006,Saridakis:2010}.\\
Furthermore, dynamical dark energy models can be extended in a huge
class of models. Among them, non-minimally coupled dark energy
models in which scalar fields coupled to the curvature terms dubbed
scalar-tensor theories have been extensively studied in the
literature. The most famous example of such theories is known as the
Brans-Dicke \cite{Brans:1961} theory in which the gravitational
constant is replaced by a scalar field $\phi$ entering into the
action as $\phi^{2}R$, $R$ is the Ricci scalar. Another well-known
example of non-minimally coupled system is provided by
$(1-\xi\phi^{2})R$ coupling in which $\xi$ is a constant measuring
the strength of non-minimal coupling
\cite{Bezrukov:2008}.\\
Moreover, due to the novel features of non-minimally coupled scalar
field system, such as allowing the phantom divide crossing and
having the cosmological scaling solutions, these models are of great
interest to the community
\cite{Sahni:2000,Sahni:2006,Linder:2008,Caldwell,Silvestri,Frieman}.
On the other hand, in 1974 Horndeski \cite{Horndeski:1974} found the
most general class of scalar-tensor theories which lead to the
second order differential equations similar to the Einstein general
relativity. The Horndeski gravity has been considered in many papers
in the context of the inflationary cosmology
\cite{Deffayet:2011,Tsujikawa:2011}. An interesting subclass of the
Horndeski theory is given by the non-minimal coupling of the scalar
field to the Gauss-Bonnet invariant in four dimensions
\cite{Antoniadis:1994,Gasperini:1997,Brustein:1998,Kawai:1998,Cartier:2000,Nojiri:2005,Calcagni:2006}.
Such a non-minimal coupling originates from the string theory and
the trace anomaly and may play an important role in cosmological
context. For example, this coupling
has been proposed to address the dark energy problem in \cite{Nojiri:2005} 
 and various aspects of accelerating cosmologies with Gauss-Bonnet correction have been
discussed in \cite{Nojiri:2006,Tsujikawa:2007,Capozziello:2013}.
Indeed, these studies yield the result that the scalar-curvature
coupling predicted by fundamental theories may become important at
current, low-curvature universe. It deserves mention that the
modifications of gravity from the Gauss-Bonnet invariant have been
often considered as the result of quantum
gravity effects \cite{Chiba:2011,Faraoni:2002,Elizalde:2004}.\\
In the present work, we will consider a model in which the scalar
field playing the role of dark energy is coupled to the Gauss-Bonnet
invariant. Here we derive constraints on the model parameters from a
combination of
available SN Ia data, as well as available BAO data and $\chi^2$ minimization technique.\\
The outline of the paper is as follows: In the next section we
present the basic formalism of our model in a flat FRW background,
along with the definition  of different cosmological parameters. We
then discuss the observational dataset and methodology in section
III. Our main results in data analysis are summarized in section IV.
Finally, section V is devoted to our conclusions.
\section{The model and cosmological background}
The model we examine in this paper is described by the following action:

\begin{equation}\label{act}
 S=\int d^{4}x\,\sqrt{-g}\,\left[\frac{R}{2\,\kappa^2}-V(\varphi)\,\sqrt{1-\partial_{\mu}\varphi\,\partial^{\mu}\varphi}
 -\eta(\varphi)\,\mathcal{G}+\mathcal{L}_{m}\right],
\end{equation}
where $g$ is the determinant of the metric tensor, $\kappa^2=8 \pi
G$, $G$ is the gravitational constant and $\mathcal{L}_{m}$ is the
matter Lagrangian density. The second term in the brackets is the
Lagrangian of tachyon field with the potential $V(\phi)$, while the
third term represents a non-minimal coupling between the scalar
field and curvature through a general function $\eta(\phi)$.
$\mathcal{G}$ is the Gauss-Bonnet invariant which is given by:
\begin{equation}\label{gauss}
\mathcal{G}=R^{2}-4\,R_{\mu\nu}\,R^{\mu\nu}+R_{\mu\nu\lambda\rho}\,R^{\mu\nu\lambda\rho},
\end{equation}
where $R$, $R_{\mu\nu}$, $R_{\mu\nu\lambda\rho}$ are the Ricci
scalar, the Ricci tensor and the Riemann tensor, respectively.\\
Notice that not only tachyon field originates from the string theory
but also the term proportional to the Gauss-Bonnet invariant
$\mathcal{G}$ is considered as a stringy correction in the action.
These are our main motivations to study the model.\\
To analyse the model it is more convenient to use the following
redefinition, as proposed in \cite{Quiros:2010} for studying the
tachyon dynamics,
\begin{equation}\label{phi}
\varphi \rightarrow \phi=\int d\varphi \sqrt{V(\varphi)}
\Longleftrightarrow
\partial \varphi=\frac{\partial \phi}{\sqrt{V(\phi)}}.
\end{equation}
Applying (\ref{phi}) in (\ref{act}) yields to our starting action as
follows:
\begin{equation}\label{action}
 S=\int d^{4}x\,\sqrt{-g}\,\left[\frac{R}{2\,\kappa^2}-V(\phi)\,\sqrt{1-\frac{\partial_{\mu}\phi\,\partial^{\mu}\phi}{V(\phi)}}
 -\eta(\phi)\mathcal{G}+\mathcal{L}_{m}\right].
\end{equation}
The variation of the action (4) with respect to the metric leads to
the following gravitational equations:
\begin{eqnarray}
R_{\mu\nu}-\frac{1}{2}g_{\mu\nu}R=\kappa^{2}\big(T_{\mu\nu}^{\phi}+T_{\mu\nu}^{GB}+T_{\mu\nu}^{m}\big),
\end{eqnarray}
where $T_{\mu\nu}^{m}$ is the usual energy-momentum tensor for the
matter, $T_{\mu\nu}^{\phi}$ corresponds to the energy-momentum
tensor of minimally coupled tachyon scalar field and
$T_{\mu\nu}^{GB}$ is the contribution of the non-minimal
Gauss-Bonnet coupling. These last two components are given by
\begin{eqnarray}
T_{\mu\nu}^{\phi}=-u\nabla_{\mu}\phi\nabla_{\nu}\phi-g_{\mu\nu}u^{-1}V(\phi)
\end{eqnarray}
and

$$T_{\mu\nu}^{GB}=4\Big(\big[\nabla_{\mu}\nabla_{\nu}\eta(\phi)\big]R-g_{\mu\nu}[\nabla_{\rho}\nabla^{\rho}\eta(\phi)\big]R
-2[\nabla^{\rho}\nabla_{\mu}\eta(\phi)\big]R_{\nu\rho}-2[\nabla^{\rho}\nabla_{\nu}\eta(\phi)\big]R_{\mu\rho}$$
\begin{eqnarray}
+2[\nabla_{\rho}\nabla^{\rho}\eta(\phi)\big]R_{\mu\nu}+2g_{\mu\nu}[\nabla^{\rho}\nabla^{\sigma}\eta(\phi)\big]R_{\rho\sigma}
-2[\nabla^{\rho}\nabla^{\sigma}\eta(\phi)\big]R_{\mu\rho\nu\sigma}\Big),
\end{eqnarray}
where
$u=\sqrt{1-\frac{\partial_{\mu}\phi\,\partial^{\mu}\phi}{V(\phi)}}$.

In the derivation of $T_{\mu\nu}^{GB}$, the properties of the
4-dimensional Gauss-Bonnet invariant have been used (see
\cite{Nojiri:2005,Farhoudi:2009} for details). The energy density
and pressure derived from these energy-momentum tensors will be
considered as effective ones and we represent them
by $\rho_{DE}$ and $p_{DE}$, respectively.\\
Now, we assume the spatially flat Friedmann-Robertson-Walker (FRW)
metric,
\begin{eqnarray}
ds^{2}=-dt^{2}+a^{2}(t)(dr^{2}+r^{2}d\Omega^{2}),
\end{eqnarray}
where $a(t)$ is the scale factor. Considering this metric in
equations (5)-(7) we obtain the following Friedmann equations:
\begin{equation}
H^{2}=\frac{\kappa^{2}}{3}\big(\rho_{DE}+\rho_{m}\big),
\end{equation}
\begin{equation}
\dot{H}=-\frac{\kappa^{2}}{2}\big(\rho_{DE}+p_{DE}+\rho_{m}+p_{m}\big),
\end{equation}
where $\rho_m$ and $p_m$ are the energy density and pressure of the
matter, $\rho_{DE}$ and $p_{DE}$ are given by
\begin{equation}
 \rho_{DE}=u\,V\left(\phi\right)+24 H^{3}\,f(\phi)\,\dot{\phi},
\end{equation}
and
\begin{equation}
 p_{DE}=-u^{-1}\,V(\phi)-8H^{2}\,\Big(f_{,\phi}\,\dot{\phi}^{2}+f(\phi)\,\ddot{\phi}\Big)-16\, H
 \,f(\phi)\,\dot{\phi}(\dot{H}+H^{2}),
\end{equation}
where $H=\frac{\dot{a}}{a}$ is the Hubble parameter, and we have
also defined $f(\phi)=\frac{d\eta}{d\phi}$, $f_{,\phi}=\frac{d f(\phi)}{d\phi}$.\\
Further, by varying the action (\ref{action}) over $\phi$ and
assuming that $\phi$ only depends on time, we obtain the equation of
motion for $\phi$, which in FRW background takes the following form
\begin{equation}
\ddot{\phi}+3\,u^{-2}\,H\,\dot{\phi}+\left(1-
\frac{3\,\dot{\phi}^{2}}{2V}\right)
V_{,\phi}+24\,H^{2}\,\left(\dot{H}+H^{2}\right)f(\phi)=0.
\end{equation}
Note that in deriving equation (13), we have used the following
expression for the Gauss-Bonnet invariant in FRW background
\begin{equation}
\mathcal{G}=24 H^2 (\dot{H}+H^2).
\end{equation}
In addition, the energy conservation equations for dark energy and
the matter are expressed in the following forms, respectively
\begin{equation}
\dot{\rho}_{DE}+3H(1+\omega_{DE})\rho_{DE}=0,
\end{equation}
and
\begin{equation}
\dot{\rho}_{m}+3H(1+\omega_{m})\rho_{m}=0,
\end{equation}
where $\omega_{DE}=\frac{p_{DE}}{\rho_{DE}}$ and
$\omega_{m}=\frac{p_{m}}{\rho_{m}}$ are the equation of state
parameters of dark energy and matter respectively. Here, we just
focus on the late-time eras, so that we can neglect the radiation
contribution and assume a pressureless fluid for the matter content
$\omega_{m}=\frac{p_{m}}{\rho_{m}}=0$. Then, the continuity equation
(16) can be easily integrated to yield
\begin{equation}\label{rho}
\rho_{m}=\rho_{m_{0}} \big(\frac{a_{0}}{a}\big)^{-3}=\rho_{m_{0}}
(1+z)^3.
\end{equation}
where $\rho_{m_{0}}$ denotes the present value of the matter energy
density and $z$ is the redshift parameter $z+1=\frac{a_{0}}{a}$. In
addition, we define the density parameters of dark energy and the
matter by  $\Omega_{DE}= (\kappa^2 \rho_{DE})/(3 H^2)$ and
$\Omega_{m}= (\kappa^2 \rho_{m})/(3 H^2)$ and here after a subscript
$"0"$ for a parameter stands for the present value of that
parameter.\\
Before closing this section, it is worthwhile to mention that since
we are going to constrain the model using observational data,
which are expressed in terms of the redshift, it is convenient to
rewrite the Friedmann equations in terms of the latter,
instead of the cosmic time.
This can be done straightforwardly by the following
replacements in equations (11) and (12),
$$\dot{H}=-HH'(1+z), \hspace{2cm}\dot{\phi}=-H(1+z)\phi',$$
\begin{equation}
\ddot{\phi}=H^{2}(1+z)\phi'+HH'(1+z)^{2}\phi'+H^{2}(1+z)^{2}\phi'',
\end{equation}
where prime denotes the derivative with respect to the redshift.
\section{Methods}
Here, we explain the methodology that we use to constrain the model
by using the recent observational datasets from Type Ia Supernova
(SNe Ia) and
Baryon Acoustic Oscillations (BAO).\\
We use the Markov-chain Monte Carlo (MCMC) method for the
minimization of $\chi^2$ to perform the statistical analysis. We
have tested the model using the publicly available codes by S.
Nesseris et a.l. (see for example
\cite{Nesseris:2013,Perivolaropoulos:2013, where the web address of the source
code can be found}) and making the necessary
changes in the case of our model. Now, we briefly explain the method for elaborating
the observational data.\\
Our study follows the likelihood $\mathcal{L}\propto \exp(-\chi^2 /2) $, where
the total $\chi^2$ for combined datasets reads:\\
\begin{equation}
\chi_{\mathrm{total}}^{2}=\chi_{\mathrm{SN}}^{2}+\chi_{\mathrm{BAO}}^{2}.
\label{eq:A.21}
\end{equation}
In the following subsections, the way by which, one can calculate
each of $\chi^2$ is described.
\subsection{Type Ia Supernova (SN Ia)}
The $\chi^2$ function for the SNe Ia is given by \cite{Perivolaropoulos:2005},\\
\begin{equation}
\chi_\mathrm{SN}^{2}=A-2\mu_{0}B+\mu_{0}^{2}C\,, \label{eq:A.5}
\end{equation}
where $A$, $B$ and $C$ are defined by
\begin{eqnarray}
&&A
=
\sum_{i}\frac{\left[\mu_{\mathrm{obs}}(z_{i})-\mu_{\mathrm{th}}(z_{i};\mu_{
0}=0)\right]^{2}}{\sigma_{i}^{2}}\,,
\quad\nonumber\\
&&B
=
\sum_{i}\frac{\mu_{\mathrm{obs}}(z_{i})-\mu_{\mathrm{th}}(z_{i};\mu_{0}=0)}
{\sigma_{i}^{2}}\,,
\quad\nonumber\\
&&C=\sum_{i}\frac{1}{\sigma_{i}^{2}}\,. \label{eq:A.6}
\end{eqnarray}
The definition of the distance modulus is
\begin{equation}
\mu_{\mathrm{th}}(z)\equiv5\log_{10}D_{L}(z)+\mu_{0}\,,
\label{eq:A.1}
\end{equation}
where $\mu_{0}\equiv42.38-5\log_{10}h$, with $h \equiv
H_{0}/100/[\mathrm{km} \, \mathrm{sec}^{-1} \, \mathrm{Mpc}^{-1}]$
\cite{Komatsu:2011} and the subscripts $"\mathrm{th}"$ and $"\mathrm{obs}"$ stand for theoretical
and the observed distance modulus. Also, the quantity $\sigma_{i}$ represents the statistical uncertainly in the distance modulus.\\
The dimensionless luminosity distance $D_{L}$ for the flat universe is given by
\begin{equation}
D_{L}(z)=\left(1+z\right)\int_{0}^{z}\frac{dz'}{E(z')}\,,
\label{eq:A.2}
\end{equation}
where
{\small{
\begin{equation}
E(z)=\frac{H(z)}{H_{0}}=
\sqrt{\Omega_{\mathrm{m}}^{(0)}\left(1+z\right)^{3}
+\Omega_{\mathrm{r}}^{(0)}\left(1+z\right)^{4}
+\Omega_{\mathrm{DE}}^{(0)}\left(1+z\right)^{3\left(1+w_{ \mathrm {
DE } } \right)} }\,. \label{eq:A.3}
\end{equation}}}
Here, $\Omega_\mathrm{r}$ is the radiation density parameter and
$\Omega_{\mathrm{r}}^{(0)}=\Omega_{\gamma}^{(0)}
\left(1+0.2271N_{\mathrm{eff}}\right)$, where
$\Omega_{\gamma}^{(0)}$ is the present fractional photon energy
density and $N_{\mathrm{eff}}=3.04$ is the effective number of
neutrino species~\cite{Komatsu:2011}.\\
Now, the minimizing of $\chi_{\mathrm{SN}}^{2}$  with respect to $\mu_{0}$ yields to
\begin{equation}
\tilde{\chi}_{\mathrm{SN}}^{2}=A-\frac{B^{2}}{C}\,.
\label{eq:A.7}
\end{equation}
In our statistical analysis we use (\ref{eq:A.7}) for SNe Ia dataset
and the Union 2.1 compilation data \cite{Suzuki:2012} of 580 data
points have been used to constrain the model parameters.
\\\\
\subsection{ Baryon Acoustic Oscillations (BAO)}
Next, we have used BAO measurement dataset to put the BAO
constraints on the model parameters. The BAO observable is the
distance ratio $d_{z}\equiv r_{s}(z_{\mathrm{d}})/D_{V}(z)$, where $r_{s}$ is the comoving sound horizon, $z_{\mathrm{d}}$ is the redshift at the drag epoch \cite{Percival:2010} and $D_{V}$ is the volume-averaged distance which is defined as follows \cite{Eisenstein:2005},\\
\begin{equation}
D_{V}(z)\equiv\left[\left(1+z\right)^{2}
D_{A}^{2}(z)\frac{z}{H(z)}\right]^{1/3} .
\label{eq:A.8}
\end{equation}
In equation (\ref{eq:A.8}) $D_{A}(z)$ is the proper angular diameter distance for the flat universe.\\
Here we have considered six BAO data points (see Table \ref{tab1}). The WiggleZ collaboration \cite{Blake:2011} has measured the baryon acoustic scale at three
different redshifts, while SDSS and 6DFGS surveys provide data at
lower redshift \cite{Percival:2010}.\\
\begin{table}[H]
\begin{center}
\begin{tabular}{| c | c | cc | ccc |}
\multicolumn{1}{c}{} & \multicolumn{1}{c}{6dF} & \multicolumn{2}{c}{SDSS}
 & \multicolumn{3}{c}{WiggleZ}  \\\hline
$z$ & 0.106 & 0.2 & 0.35 & 0.44 & 0.6 & 0.73  \\\hline
$d_z$ & 0.336 & 0.1905 & 0.1097 & 0.0916 & 0.0726 & 0.0592  \\\hline
$\Delta d_z$ & 0.015 & 0.0061 & 0.0036 & 0.0071 & 0.0034 & 0.0032 \\\hline
 \end{tabular}
 \caption{The BAO data used in our analysis.\label{tab1}}
\end{center}
\end{table}
The $\chi^2$ function of the BAO data is defined as,\\
%
\begin{equation}
\chi_{\mathrm{BAO}}^{2}=
\left(x_{i,\mathrm{BAO}}^{\mathrm{th}}-x_{i,\mathrm{BAO}}^{\mathrm{obs}}
\right) \left(C_{\mathrm{BAO}}^{-1}\right)_{ij}
\left(x_{j,\mathrm{BAO}}^{\mathrm{th}}-x_{j,\mathrm{BAO}}^{\mathrm{obs}}
\right), \label{eq:A.14}
\end{equation}
%
where the indices $i,j$ are in growing order in $z$, as in Table
\ref{tab1} and  $C_{\mathrm{BAO}}^{-1}$ can be obtained by the
covariance data \cite{Blake:2011} in terms of $d_z$ as follows:
%
\begin{equation}
C_{\mathrm{BAO}}^{-1}=\left(
\begin{array}{cccccc}
4444&0&0&0&0&0\\
0&30318&-17312&0&0&0\\
0&-17312&87046&0&0&0\\
0&0&0&23857&-22747&10586\\
0&0&0&-22747&128729&-59907\\
0&0&0&10586&-59907&125536
\end{array}\right)\,.
\label{eq:A.13}
\end{equation}
%
One can now obtain the best fit values of the model parameters by
minimizing $\chi_{\mathrm{total}}^{2}$ in equation (\ref{eq:A.21}).
\section{Observational constraints on the model parameters}
Following the $\chi^2$ analysis ( as presented in the previous
section), in this section, we obtain the constraints on the free
parameters of the model.\\
Two important functions in our analysis are the function $f(\phi)$
and the scalar field potential $V(\phi)$. We will consider power-law
and exponential forms for $f(\phi)$ and $V(\phi)$ and thus our study
is categorized into the following four different cases:\\
Case I: Exponential $f(\phi)$ and Power-law $V(\phi)$ $$f(\phi)
\propto e^{\alpha \phi} ,  V(\phi) \propto \phi^\beta$$ Case II:
Power-law $f(\phi)$ and Exponential $V(\phi)$  $$f(\phi)
\propto\phi^\alpha , V(\phi) \propto e^{\beta \phi}$$ Case III:
Power-law $f(\phi)$ and $V(\phi)$   $$f(\phi) \propto  \phi^\alpha,
V(\phi) \propto \phi^\beta$$ Case IV: Exponential $f(\phi)$ and
$V(\phi)$  $$f(\phi) \propto e^{\alpha \phi}, V(\phi) \propto
e^{\beta \phi},$$
where $\alpha$ and $\beta$ are the free parameters which will be constrained using the data. \\
In the present work, we identify the parameters of the model as the
parameters $\alpha$ and $\beta$, the present matter density
parameter $\Omega_{m_{0}}$ and dark energy equation of state
parameter $\omega_{DE}$ . Thus, the model parameters are
($\alpha,~\beta,~ \Omega_{m_{0}}
,~\omega_{DE}$ ).\\
Now, performing the combined ($SN Ia + BAO$) analysis using MCMC
method for the cases I-IV, yields to the constraint results as what
summarized in Table 2.
\begin{table}[H]
\begin{center}
\begin{tabular}{| c | c | c | c | c | c |}
\hline $model$&$\chi_{min}^2$&$\Omega_{m_{0}}$&$\omega_{DE}$&$\alpha$&$\beta$\\
\hline
Case I &$594.216$&$0.26\pm 0.006$&$-1.099\pm 0.002$&$-0.995\pm 0.012$&$0.023\pm 0.011$\\
\hline
Case II&$663.505$&$0.26\pm 0.003$&$-1.199\pm 0.003$&$-2.998\pm 0.025$&$0.077\pm 0.013$\\
\hline
Case III&$601.833$&$0.26\pm 0.005$&$-1.199\pm 0.004$&$-1.998\pm 0.025$&$0.23\pm 0.016$\\
\hline
Case IV&$595.747$&$0.26\pm 0.028$&$-1.25\pm 0.06$&$-1.96\pm 0.25$&$0.43\pm 0.081$\\
\hline
\end{tabular}
\caption{The value of $\chi_{min}^2$ and the best fit values of the
model parameters $\Omega_{m_{0}}$, $\omega_{DE}$, $\alpha$ and
$\beta$ for the cases I-IV.} \label{tab2}
\end{center}
\end{table}
In this table, the reader may see a compact presentation of the best
fit values of the model parameters as well as $\chi^{2}_{min}$ for each case, separately. \\
Furthermore, in Figures $1-4$ we present the $1\sigma, 2\sigma$ and
$3\sigma$ confidence level contour plots for several combinations of
the model parameters as well as their likelihood analysis for the
cases I-IV, respectively. Additionally, in figure \ref{wz}, using
the same combined analysis SN Ia + BAO, we have shown the
qualitative evolution of the dark energy equation of state
parameter. Figure \ref{cd} shows the Hubble diagram for 580 SN Ia
from (Union 2.1) sample. The curves represent the distance modulus
predicted by the four cases I-IV in our model.\\
Further, the case with the lowest value of  $\chi^{2}_{min}$ is the
case I and as it is clear from Table \ref{tab2} the cases with the
exponential coupling function $f(\phi)$ (cases I \& IV) have a lower
$\chi^{2}_{min}$ than the cases with power-law $f(\phi)$ (cases II \& III).\\
The joint analysis on cases I-IV shows that the best fit values of
the dark energy equation of state parameter, exhibit phantom
behavior, although very close to the cosmological constant boundary.
As one see from Table \ref{tab2}, in case I, we have the nearest
value of the equation of state parameter to the cosmological
constant ($\omega_{\Lambda}=-1$) and in case IV, the phantom
character of the current dark energy equation of state is the
clearest one. Notice that the quintessence behavior of $\omega_{DE}$
is excluded in all cases of Table \ref{tab2}. Further, from Figure
\ref{wz}, it is clear that the transition from quintessence phase
($\omega_{DE} >- 1$) to the phantom phase ($\omega_{DE} <- 1$) or
the so-called
phantom divide line crossing, occurs in all four cases, which is in agreement
with observational results \cite{Zhao:2005,Caldwell:2005,Feng:2005}.\\
From the constraints on $\alpha$ and $\beta$ as shown in Table \ref{tab2}, it is clear that the combination of
SN Ia + BAO data favors negative values for $\alpha$ and positive values for $\beta$ in cases I-IV.\\
Between the best-fit values of $\alpha$, the case II i.e. when the
coupling function is in power-law form, has the minimum value, while
the case I in which $f(\phi)$ is exponential has the maximum value
of $\alpha$. On the other hand, the best-fit value of the free
parameter $\beta$, has its minimum and maximum  values for the cases
I and IV, that is for power-law and exponential potentials,
respectively.\\
It deserves mention here that the values of the dark matter density
parameter at present $\Omega_{m_{0}}$ for all four cases
are very close to the desired value in cosmology.\\

\begin{figure}[H]
    \centering
      \includegraphics[width=120mm]{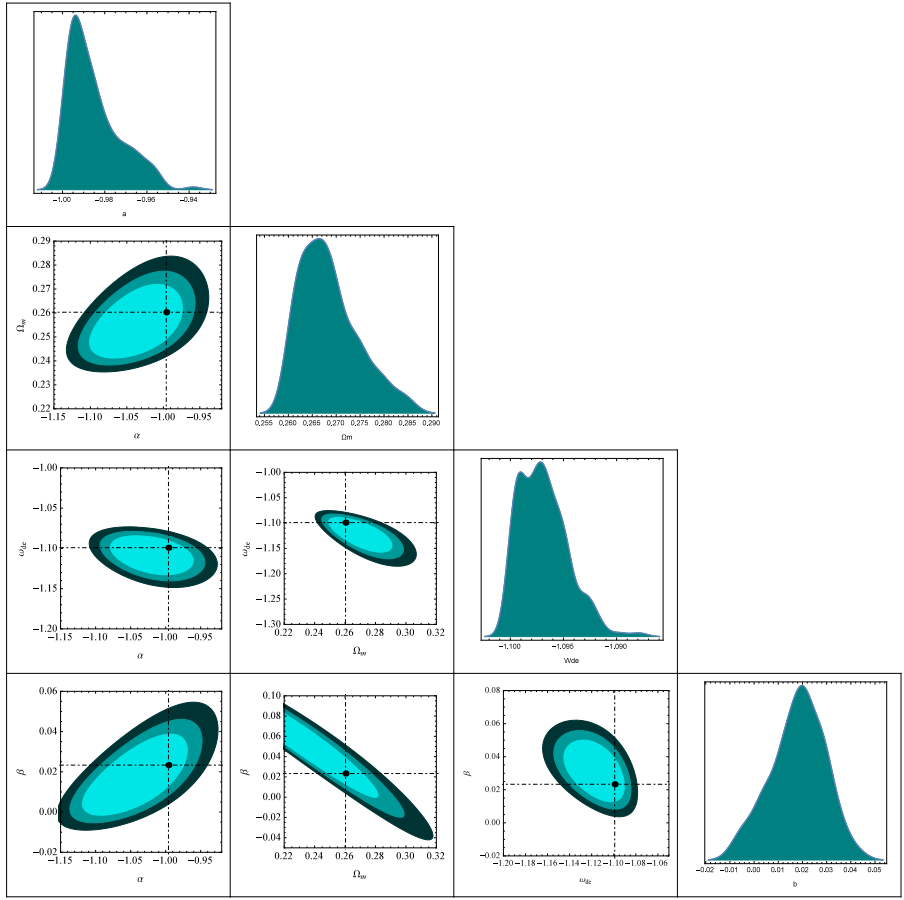}
        \caption{$1\sigma$ (68.3\%), $2\sigma$ (95.4\%) ans $3\sigma$ (99.7\%) confidence level contour plots for different combinations of the model
        parameters with also 1-dimensional posterior distributions in the case I for combined observational dataset from SN Ia + BAO. The black dot in each contour plot represents the best fit values of the corresponding pair. }
    \label{fexp}
\end{figure}

\begin{figure}[H]
    \centering
      \includegraphics[width=120mm]{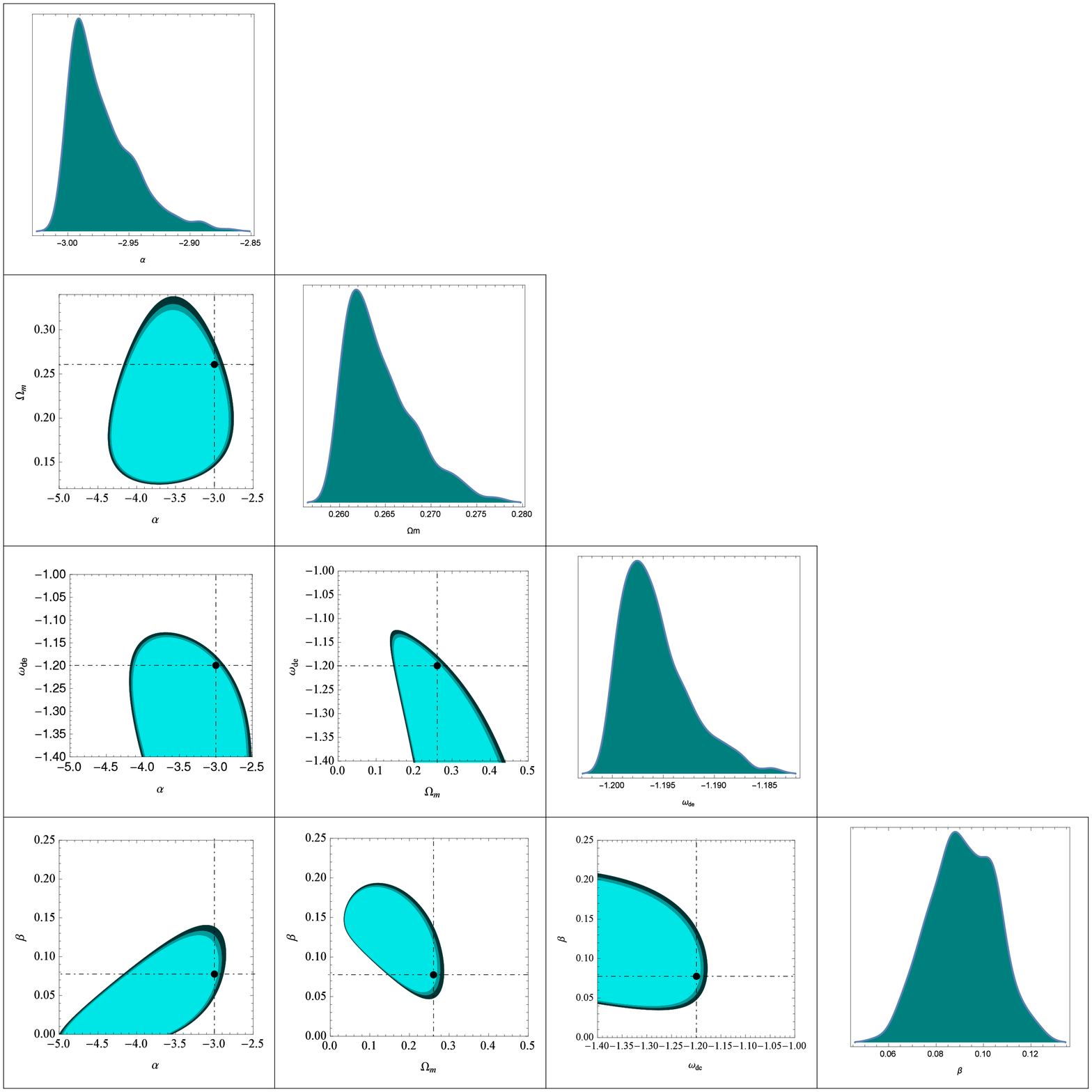}
        \caption{$1\sigma$ (68.3\%), $2\sigma$ (95.4\%) ans $3\sigma$ (99.7\%) confidence level contour plots for different combinations of the model
        parameters with also 1-dimensional posterior distributions in the case II for  combined observational dataset from SN Ia + BAO. The black dot in each contour plot represents the best fit values of the corresponding pair. }
    \label{vexp}
\end{figure}

\begin{figure}[H]
    \centering
      \includegraphics[width=120mm]{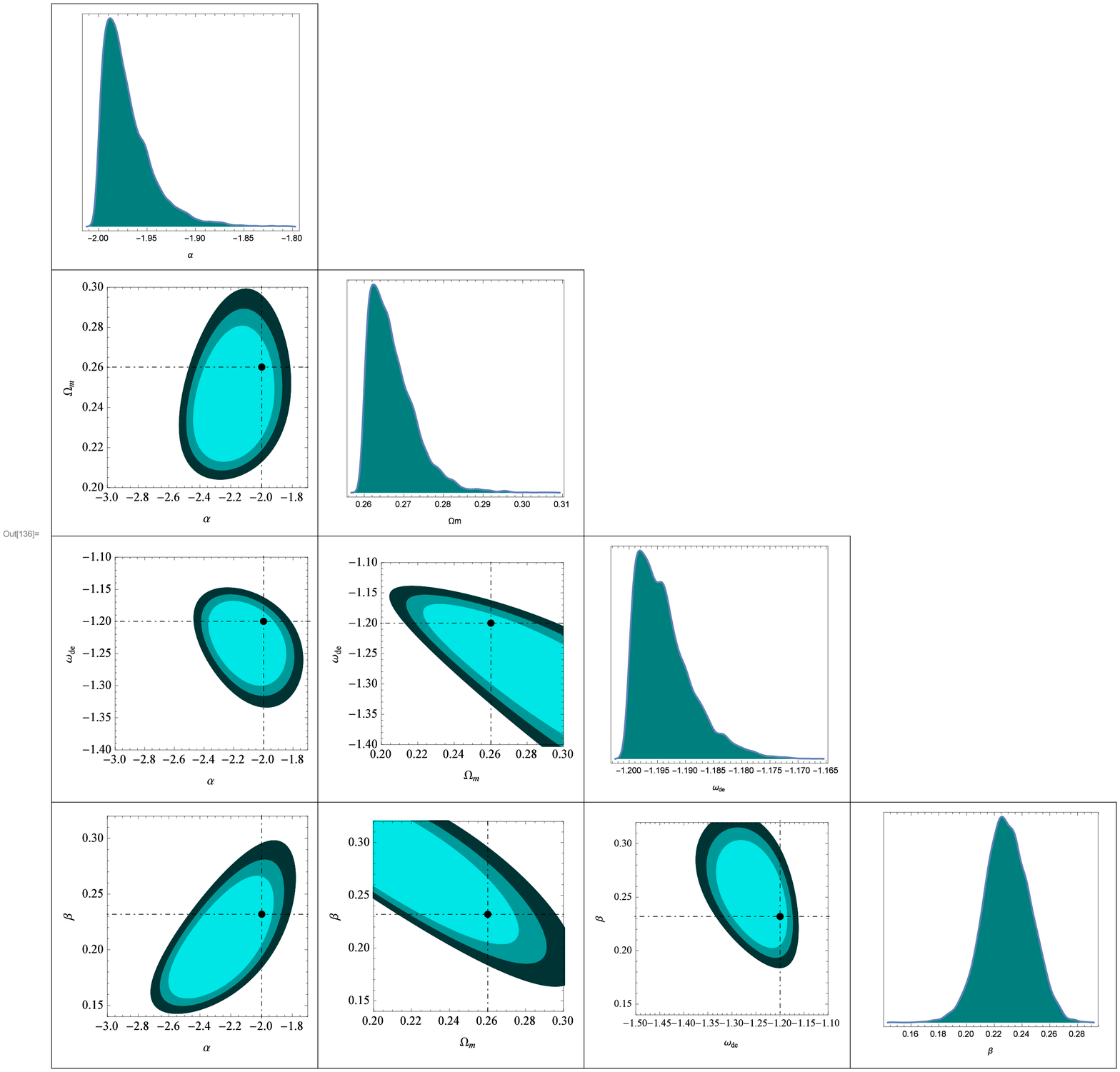}
        \caption{$1\sigma$ (68.3\%), $2\sigma$ (95.4\%) ans $3\sigma$ (99.7\%) confidence level contour plots for different combinations of the model
        parameters with also 1-dimensional posterior distributions in the case III for combined observational dataset from SN Ia + BAO. The black dot in each contour plot represents the best fit values of the corresponding pair. }
    \label{notexp}
\end{figure}

\begin{figure}[H]
    \centering
      \includegraphics[width=120mm]{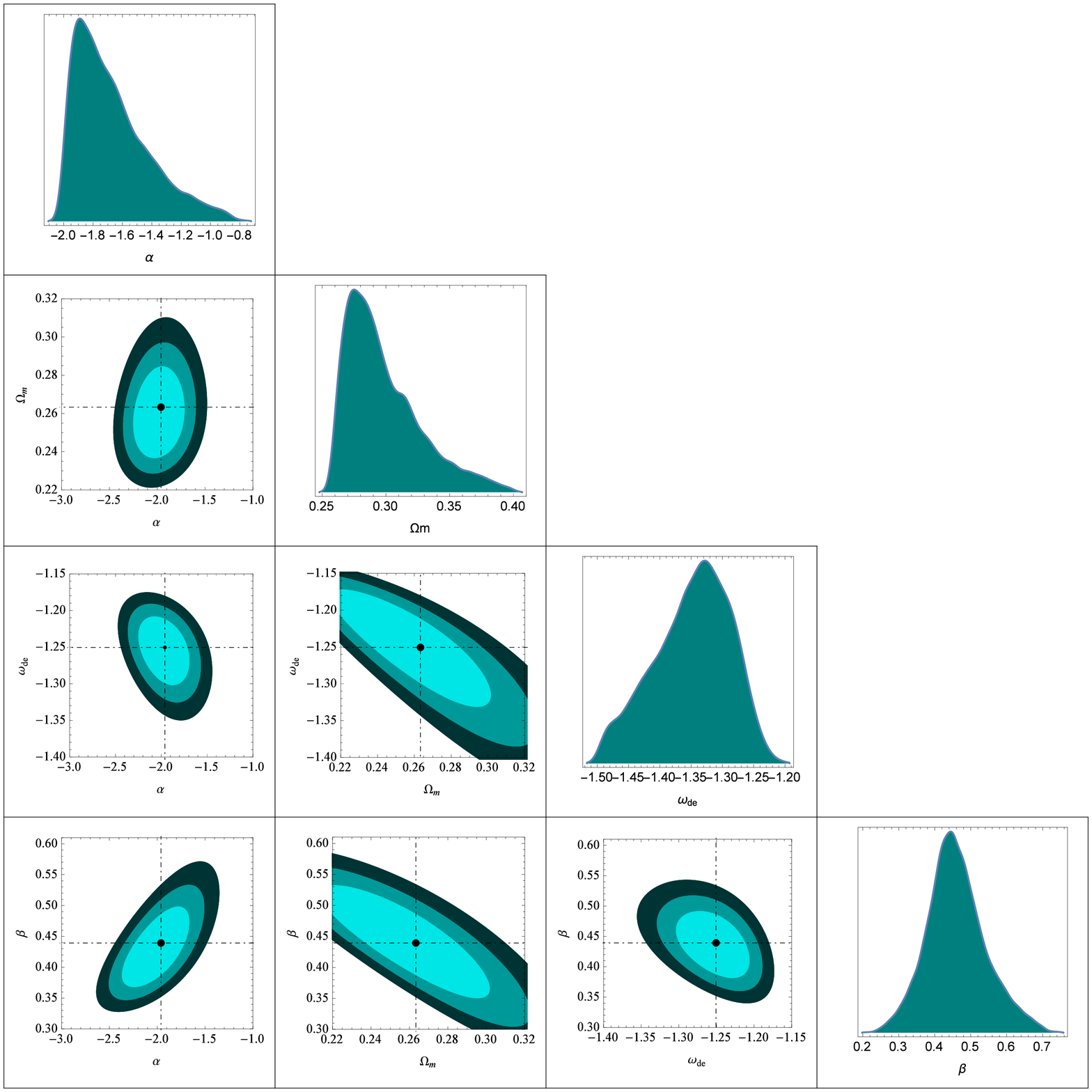}
        \caption{$1\sigma$ (68.3\%), $2\sigma$ (95.4\%) ans $3\sigma$ (99.7\%) confidence level contour plots for different combinations of the model
        parameters with also 1-dimensional posterior distributions in the case IV for combined observational dataset from SN Ia + BAO.
        The black dot in each contour plot represents the best fit values of the corresponding pair. }
    \label{fvexp}
\end{figure}
\begin{figure}[H]
    \centering
      \includegraphics[width=100mm]{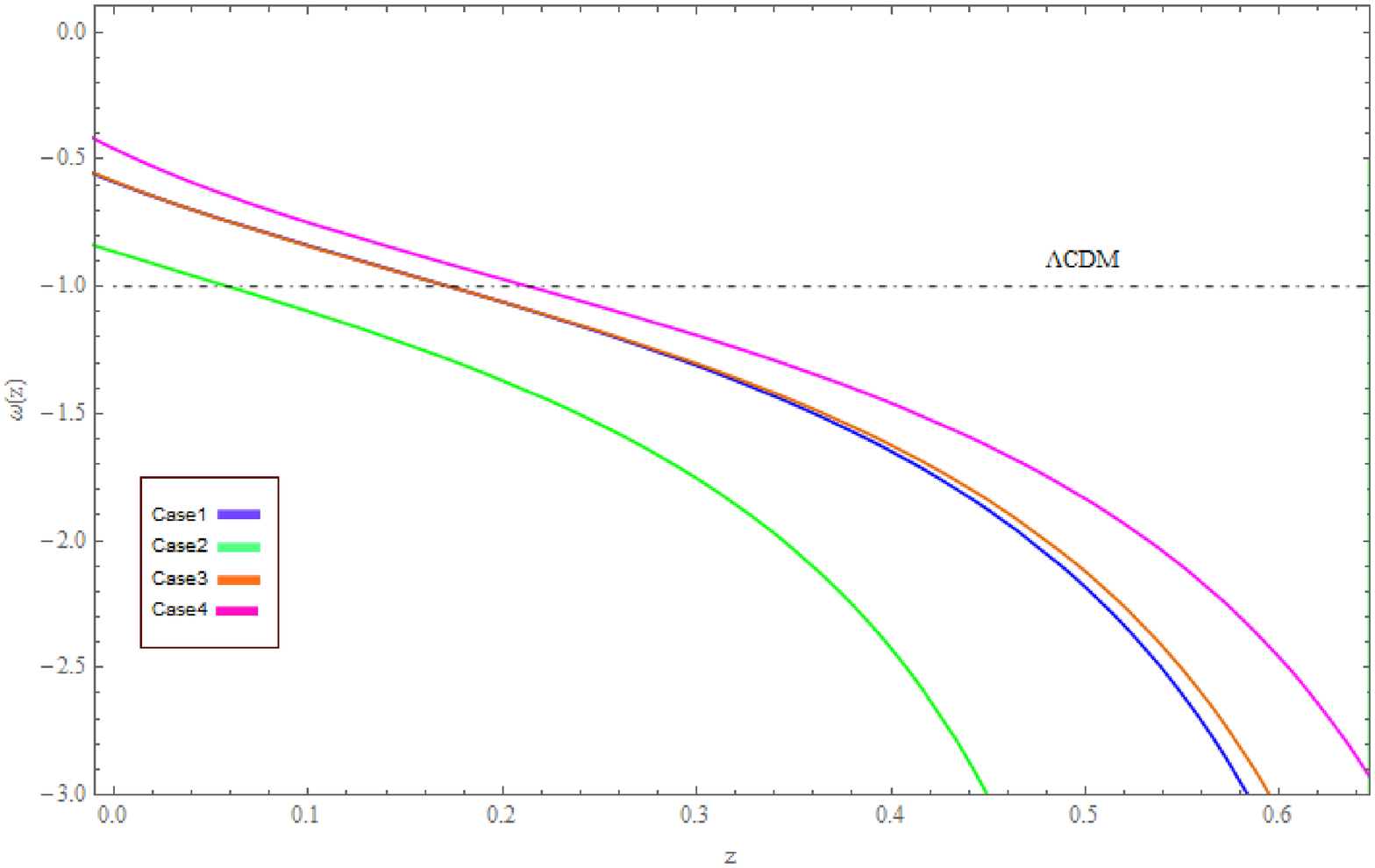}
        \caption{The evolution of the dark energy equation of state parameter, for the best fit values of ($\alpha , \beta$) that arises from the analysis of SN Ia + BAO datasets, for the cases I (blue), II(green), III (orange) and IV (pink).}
    \label{wz}
\end{figure}
\begin{figure}[H]
    \centering
      \includegraphics[width=100mm]{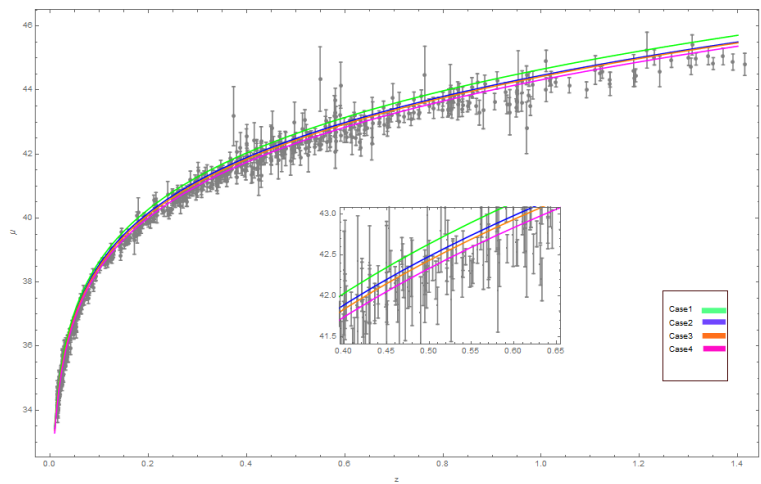}
        \caption{The Hubble diagram for 580 data of SN Ia from Union 2.1 sample \cite{Suzuki:2012}.The curves correspond to the
        distance modulus predicted by the four cases I-IV with the best-fit values coming from the joint
        analysis of SN Ia + BAO as presented in Table \ref{tab2}. }
    \label{cd}
\end{figure}

\section{Conclusion}
In this paper, we have focused on the analysis of a non-minimally
coupled scalar field theory in which the scalar field is considered
as a candidate of dark energy. In this model a tachyon field is
non-minimally coupled to the Gauss-Bonnet invariant, via a general
coupling function $f(\phi)$, as in action (4). The cosmological
evolution of the model is studied by assuming a flat FRW universe.
Then, we placed constraints on the free parameters of the model by
performing a joint statistical analysis using the recent
cosmological data from SN Ia and BAO measurements. We have
considered the exponential and power-law forms for the scalar field
potential, as well as the non-minimal coupling function. Then, we
have obtained the best fit values of the free parameters $\alpha$
and $\beta$ in the potential and the coupling function, respectively.
The equation of state parameter of dark energy, $\omega_{DE}$ and
the present value of the matter density $\Omega_{m_{0}}$ have also
been fitted.\\
According to the contents of Table 2, where our results are
summarized, the joint analysis of SN Ia $+$ BAO, favors the negative
values for $\alpha$ and positive values for $\beta$. In addition, by
constraining with the datasets of SN Ia and BAO, we found that
$\omega_{DE}<-1$, for all cases, which means our universe slightly
biases towards phantom behavior while the values of
$\Omega_{m_{0}}$, i.e. the present-day dark matter density, are very close
to the desired value $\Omega_{m_{0}}\simeq 0.27$. Finally, using the
best fit values of the model parameters in Table 2, we have evolved
the equation of state parameter, $\omega_{DE}$ for all cases I-IV in
Figure 5. The so-called phantom divide line crossing phenomenon has
been clearly depicted in this figure. All in all, according to our
analysis, a theory with a non-minimal coupling between the tachyon scalar
field and the Gauss-Bonnet invariant is in agreement with
cosmological observations and can be considered
as a good candidate for dark energy.\\

\end{document}